\documentclass[amssymb,11pt]{article}
\usepackage[pdftex,colorlinks=true,urlcolor=black,filecolor=black,linkcolor=black,
            pdftitle={Quadratic field equations on the brane.},
            pdfauthor={Mark D. Roberts},
            pdfsubject={relativity, five dimensions},
            pdfkeywords={Gauss,Bonnet,Bach,Pauli},
            pagebackref,pdfpagemode=None,bookmarksopen=true]{hyperref}
\usepackage{amssymb}
\newcommand{\bc}{\begin{center}}
\newcommand{\ec}{\end{center}}
\newcommand{\bi}{\begin{itemize}}     
\newcommand{\ei}{\end{itemize}}
\newcommand{\bd}{\begin{description}} 
\newcommand{\ed}{\end{description}}
\newcommand{\bn}{\begin{enumerate}}   
\newcommand{\en}{\end{enumerate}}
\newcommand{\be}{\begin{equation}}
\newcommand{\ee}{\end{equation}}
\newcommand{\ber}{\begin{eqnarray}}
\newcommand{\ear}{\end{eqnarray}}
\newcommand{\ba}{\begin{array}}
\newcommand{\ea}{\end{array}}
\newcommand{\bv}{\begin{verbatim}}
\newcommand{\ev}{\end{verbatim}}

\newcommand{\bx}{\Box}

\newcommand{\de}{\delta}

\newcommand{\fr}{\frac}

\newcommand{\n}{\nonumber\\}

\newcommand{\st}{\stackrel}

\begin{document}
\title{Quadratic field equations on the brane.}
\author{
\href{http://www.violinist.com/directory/bio.cfm?member=robemark}
{Mark D. Roberts},\\
54 Grantley Avenue,  Wonersh Park,  GU5 0QN,  UK\\
mdr@ihes.fr
}
\date{$21^{st}$ of March 2010}
\maketitle
\begin{abstract}
It is shown that four dimensional vacuum Einstein solutions simply embedded
in five dimensions obey the Gauss-Bonnet-Einstein field equations:
$G_{ab}+\alpha GB_{ab}+\delta^{55}_{ab}\alpha\exp(-2\chi/\sqrt{\alpha})GB_4=0$
and the Pauli-Einstein equations
$G_{ab}-3\alpha P_{ab}/5=0$,
and the Bach-Einstein equations $B_{ab}=0$.
General equations are calculated for which these and similar results follow.
It is briefly argued that such field equations could be significant on large distance scales.
\end{abstract}
\section{Introduction.}\label{intro}
Two frequent variants on Einstein's equations are equations involving
quadratic lagrangians and those involving higher dimensions,
and there is the possibility of these variants describing dynamics on large distance scales.
This paper studies the relationship between solutions to the four
dimensional Einstein equations and these more complex equations.
The curvature tensors corresponding to quadratic lagrangians can be split into three parts:
the Gauss-Bonnet tensor $GB_{ab}$ which vanishes in four dimensions,
the Pauli tensor $P_{ab}$ corresponding to the $R^2$ part of the lagrangian,
and the Bach tensor $B_{ab}$ corresponding to the Weyl tensor squared part of the lagrangian.
Here the five dimensional line element
\be
ds_5^2=\exp(\fr{\chi}{\sqrt{\alpha}})ds_4^2+d\chi^2,
\label{le}
\ee
where $ds_4^2$ is a given four dimensional solution is considered.
The quadratic tensors corresponding to this line element are worked out directly,
although perhaps they could be worked out using existing formalisms such as the ADM
\cite{ADM62} or the Gauss-Coddacci equations \cite{HE}.

The type of questions the resulting equations allow to be addresses are:
when are ''black holes'', such as the four dimensional Schwarzschild solution,
solutions to five dimensional equations,  for previous work see \cite{DMPR,MM,leon,DG};
when are four dimensional radiating solutions solutions to five dimensional equations,
for previous work see \cite{mdrfi,DG};
and when are other four dimensional solutions,  such as the Kasner solution,
solutions to five dimensional equations \cite{mdrkb}.
Having Einstein-Gauss-Bonnet equations obeyed in the bulk leads to
better approximations to Newton's law on the brane \cite{DS,KL}.
The conventions used are early latin indices $a,b,\dots$ for five dimensional quantities,
middle latin indices for four dimensional quantities $i,j,\dots$,
signature $-++++$.
\section{Equations.}\label{equations}
For the line element \ref{le} the five dimensional metric and its determinant are
\be
\st{5}{g}_{ij}=\exp(\fr{\chi}{\sqrt{\alpha}})\st{4}{g}_{ij},
\st{5}{g^{ij}}=\exp(-\fr{\chi}{\sqrt{\alpha}})\st{4}{g^{ij}},
g_{55}=1,
g^{55}=1,
\st{5}{g}=\exp(\fr{4\chi}{\sqrt{\alpha}})\st{4}{g}.
\ee
The five dimensional christoffel symbols are
\be
\st{5}{\{^i_{jk}\}}=\st{4}{\{^i_{jk}\}},
\{^5_{ij}\}=-\fr{1}{2\sqrt{\alpha}}\exp(\fr{\chi}{\sqrt{\alpha}})\st{4}{g_{ij}},
\{^i_{j5}\}=\fr{1}{2\alpha}\de^i_j,
\{^5_{55}\}=\{^5_{5i}\}=\{^i_{55}\}=0.
\ee
The five dimensional Riemann tensor can be expressed in terms of four dimensional quantities
\ber
&&R^5_{.i5j}=-\fr{1}{4\alpha}\exp(\fr{\chi}{\sqrt{\alpha}})\st{4}{g}_{ij},~~~
R^5_{.ijk}=0,\\
&&\st{5}{R^i}_{.jkl}=\st{4}{R^i}_{.jkl}
+\fr{1}{4\alpha}\exp(\fr{\chi}{\sqrt{\alpha}})
\left(-\de^i_k\st{4}{g}_{lj}+\de^{i_l}\st{4}{g}_{kj}\right).
\nonumber
\label{rie5}
\ear
The five dimensional Ricci tensor,  defined by $R_{bd}=R^c_{.bcd}$,
can be expressed in terms of four dimensional quantities
\be
R_{55}=-\fr{1}{\alpha},~~~
R_{5i}=0,~~~
\st{5}{R}_{ij}=\st{4}{R}_{ij}-\fr{1}{\alpha}\exp(\fr{\chi}{\sqrt{\alpha}})\st{4}{g}_{ij},
\label{Ricci5}
\ee
the Ricci scalar is
\be
\st{5}{R}=-\fr{5}{\alpha}+\exp(-\fr{\chi}{\sqrt{\alpha}})\st{4}{R}.
\label{Rs5}
\ee
The five dimensional Einstein tensor can be expressed in terms of four dimensional quantities
\be
G_{55}=\fr{3}{2\alpha}-\fr{1}{2}\exp(-\fr{\chi}{\sqrt{\alpha}})\st{4}{R},~~~
\st{5}{G}_{ij}=\fr{3}{2\alpha}\exp(\fr{\chi}{\sqrt{\alpha}})\st{4}{g}_{ij}
+\st{4}{G}_{ij},~~~
\st{5}{G}=-\fr{3}{2}\st{5}{R}.
\label{Ein5}
\ee
The Weyl tensor \cite{HE} p.41 is defined as
\ber
C_{abcd}&\equiv&R_{abcd}+\fr{R}{(d-1)(d-2)}(g_{ac}g_{bd}-g_{ad}g_{cb})\n
&&+\fr{1}{(d-2)}(g_{ad}R_{cb}-g_{ac}R_{db}+g_{bc}R_{da}-g_{bd}R_{ca}),
\label{weyl}
\ear
expressing the five dimensional Weyl tensor in term of four dimensional quantities
\ber
&&C^5_{.i5j}=-\fr{1}{3}\st{4}{R}_{ij}+\fr{1}{12}\st{4}{g}_{ij}\st{4}{R},~~~~~
C^5_{.ijk}=0,\\
&&\st{5}{C}^i_{.jkl}=\st{4}{C}^i_{.jkl}
-\fr{1}{6}
(g^i_l\st{4}{R}_{kj}-\de^i_k\st{4}{R}_{lj}+g_{jk}\st{4}{R}^i_{.l}-g_{lj}\st{4}{R}^i_{.k})
-\fr{1}{12}\st{4}{R}(\de^i_kg_{jl}-\de^i_lg_{kj}).
\nonumber
\ear
The invariant quadratic products in arbitrary dimension $d$ obey
\be
Wy=Rm-\fr{4}{(d-2)}Rc+\fr{2}{(d-1)(d-2)}R^2,
\ee
where
\be
Rc\equiv R_{ab}R^{ab},~~~
Rm\equiv R_{abcd}R^{abcd},~~~
Wy\equiv C_{abcd}C^{abcd}.
\ee
Expressing these four dimensional invariants in terms of five dimensional quantities
\ber
\st{5}{R}^2&=&\fr{25}{\alpha^2}-\fr{10}{\alpha}\exp(-\fr{\chi}{\sqrt{\alpha}})\st{4}{R}
+\exp(-2\fr{\chi}{\sqrt{\alpha}})\st{4}{R}^2,\\
\st{5}{Rc}&=&\fr{5}{\alpha^2}-\fr{2}{\alpha}\exp(-\fr{\chi}{\sqrt{\alpha}})\st{4}{R}
+\exp(-2\fr{\chi}{\sqrt{\alpha}})\st{4}{Rc},\n
\st{5}{Rm}&=&\fr{5}{2\alpha^2}-\fr{1}{\alpha}\exp(-\fr{\chi}{\sqrt{\alpha}})\st{4}{R}
+\exp(-2\fr{\chi}{\sqrt{\alpha}})\st{4}{Rm},\n
\st{5}{Wy}&=&\exp(-2\fr{\chi}{\sqrt{\alpha}})
\left[\st{4}{Wy}+\fr{2}{3}\st{4}{Rc}-\fr{1}{6}\st{4}{R}^2\right],
\nonumber
\label{inv5}
\ear
The Gauss-Bonnet combination is defined by
\be
GB\equiv Rm-4Rc+R^2,
\label{gbc}
\ee
expressing the five dimensional object in terms of four dimensional objects
\be
\st{5}{GB}=\fr{15}{2\alpha^2}-\fr{3}{\alpha}\exp(-\fr{\chi}{\sqrt{\alpha}})\st{4}{R}
+\exp(-2\fr{\chi}{\sqrt{\alpha}})\st{4}{GB}.
\label{gbc5}
\ee
The Gauss-Bonnet tensor is defined by
\be
GB_{a b}\equiv4R_{acde}R_b^{~cde}-8R_{cd}R^{c~d}_{.a.b}-8R_{ac}R^c_{.b}+4RR_{ab}-g_{ab}GB,
\label{gbt}
\ee
expressing this tensor in terms of four dimensional objects
\ber
&&\st{5}{GB}_{55}=-\fr{3}{2\alpha^2}+\fr{1}{\alpha}\exp(-\fr{\chi}{\sqrt{\alpha}})\st{4}{R}-\exp(-2\fr{\chi}{\sqrt{\alpha}})\st{4}{GB},\\
&&\st{5}{GB}_{ij}=-\fr{3}{2\alpha^2}\exp(\fr{\chi}{\sqrt{\alpha}})g_{ij}
-\fr{2}{\alpha}\st{4}{G}_{ij}
+\exp(-\fr{\chi}{\sqrt{\alpha}})\st{4}{GB}_{ij}
\nonumber
\label{gb5}
\ear
The Pauli tensor is defined by
\be
P_{ab}\equiv2R_{;ab}-2RR_{ab}+g_{ab}(\fr{1}{2}R^2-2\bx R),
\label{pauli}
\ee
expressing the Pauli tensor in terms of four dimensional objects
\ber
&&P_{55}=\fr{5}{2\alpha^2}
+\fr{1}{\alpha}\exp(-\fr{\chi}{\sqrt{\alpha}})\st{4}{R}
+\exp(-2\fr{\chi}{\sqrt{\alpha}})(\fr{1}{2}\st{4}{R}-2\st{4}{\bx}\st{4}{R}),\n
&&P_{5i}=-\fr{3}{\sqrt{\alpha}}\exp(-\fr{\chi}{\sqrt{\alpha}})R_{,i},\\
&&\st{5}{P}_{ij}=\exp(-\fr{\chi}{\sqrt{\alpha}})\st{4}{P}_{ij}
+\fr{5}{2\alpha^2}\exp(\fr{\chi}{\sqrt{\alpha}})g_{ij}
-\fr{2}{\alpha}\st{4}{R}g_{ij}+\fr{10}{\alpha}\st{4}{R}_{ij},
\nonumber
\ear
note the non-vanishing off diagonal $P_{5i}$ term.
The Bach tensor is defined by
\be
B_{ab}\equiv2C_{a..b}^{~cd}R_{cd}+4C_{a..b;cd}^{~cd},
\label{bachdef}
\ee
and has trace $B^a_{.a}=0$,
expressing the Bach tensor in terms of four dimensional objects
\ber
&&B_{55}=\exp(-2\fr{\chi}{\sqrt{\alpha}})
\left\{+\fr{2}{3}Rc-\fr{1}{6}R^2+\fr{4}{3}R^{ij}_{..;ij}-\fr{1}{3}\st{4}{\bx}R\right\},\\
&&\st{5}{B}_{ij}=\exp(-2\fr{\chi}{\sqrt{\alpha}})\st{4}{B}_{ij}
-\fr{4}{\alpha}\exp(-\fr{\chi}{\sqrt{\alpha}})C^5_{.i5j}
+\exp(-2\fr{\chi}{\sqrt{\alpha}})\left\{H_{ij}+X_{ij}\right\},\nonumber
\label{B5ij}
\ear
where the higher order terms $H_{ij}$ and the cross terms $X_{ij}$ are given by
\ber
&&3H_{ij}\equiv-2{\bx}R_{ij}+2R^{~k}_{j.;ki}+2R^{~k}_{i.;kj}-R_{;ij}
+g_{ij}(-2R^{kl}_{..;kl}+\st{4}{\bx}R),\n
&&3X_{ij}\equiv-8S_{ij}-\fr{1}{6}RR_{ij},
\label{HX}
\ear
up to coupling constant, the sms \cite{SMS} product tensor $S_{ij}$ can be
expressed in terms of the Ricci tensor rather than the stress and is
\be
S_{ij}=\fr{1}{6}RR_{ij}-\fr{1}{4}R_{ik}R_j^{~k}+g_{ij}(\fr{1}{8}Rc-\fr{1}{16}R^2).
\ee
The cross terms in the Bach tensor are nearly the same as for the sms tensor,
there being an additional $RR_{ij}$ term.
\section{Solutions}\label{solutions}
The {\it first} set of solutions is that the vacuum-Bach equations are obeyed when
\be
\st{4}{R}_{ij}=\st{4}{\lambda}\st{4}{g}_{ij}.
\label{l4}
\ee
In general this is not the same solution as
\be
\st{5}{R}_{ab}=\st{5}{\lambda}\st{5}{g}_{ab},
\label{l5}
\ee
substituting (\ref{l5}) into (\ref{bachdef}) the Bach tensor always vanishes,
substituting (\ref{l4}) into (\ref{B5ij}) gives vanishing Bach tensor,
however substituting (\ref{l4}) into the Ricci tensor (\ref{Ricci5})
gives
\be
R_{55}=-\fr{1}{\alpha},~~~
\st{5}{R}_{ij}=\left(\st{4}{\lambda}\exp(-\fr{\chi}{\sqrt{\alpha}})-\fr{1}{\alpha}\right)\st{5}{g}_{ij},
\ee
which is more general than (\ref{l5}) and reduces to (\ref{l5}) when
$\st{4}{\lambda}=0,~-1/\alpha=\st{5}{\lambda}$.

The {\it second} set of solutions are that
for $\st{4}{R}_{ij}=0$, which implies $\st{4}{P}_{ij}=0$,
\be
G_{ab}+\alpha\beta GB_{ab}
+\fr{3}{5}\alpha(\beta-1)P_{ab}
+\delta^{55}_{ab}\alpha\exp(-2\chi/\sqrt{\alpha})\st{4}{GB}=0,
\label{Ff}
\ee
where $\beta$ is an arbitrary constant,
the solutions in the abstract are obtained when $\beta=0,1$.
That for $\beta=0,1$ the field equations are obeyed
by the four dimensional Schwarzschild and Kasner
solutions was noted in \cite{mdrfi,mdrkb}.
Despite the arbitrary constant $\beta$
there does not appeared to be any other simple solutions.
For example the ansatz (\ref{l4}) applied to the Bach and Pauli equations does not
seem to lead to generalizations of (\ref{Ff}),
similarly for the ansatz $\st{4}{R}=0$.
For $\st{4}{R}_{ij}$ coupled to scalar fields,  vector fields and fluids
there is not much simplification of the quadratic tensors.
The above does not indicate why the Gauss-Bonnet tensor vanishes for the
first and second five dimensional scalar solutions of \cite{mdrfi},
because for these spacetimes the geometry is different from (\ref{le}).
\section{Conclusion.}\label{conclusion}
On length scales of about that of a stellar cluster and larger it seems that the
perfect fluid Einstein equations no longer work
because they are not able to explain constant galactic rotation curves \cite{mdrhs};
so the question arises as to what field equations determine dynamics on large scales.
From string theory it seems that the most likely variant of general relativity at large
scales involve the Gauss-Bonnet tensor \cite{zwiebach}.
The Gauss-Bonnet combination in $d=5$ Randall-Sundrum setup has been explicitly shown
to only give rise to short-distance corrections to the brane Newton potential \cite{DS,KL};
on the other hand brane gravity in an infinite transverse volume
$d=5$ Einstein-Hilbert-Gauss-Bonnet theory
might display a large-distance non-conventional behavior leading to different laws of attraction.
Four dimensional vacuum Einstein equations certainly seem to hold on middle length scales,
as it is from them that contact is made with Newtonian theory.
These two facts suggest that field equations such as (\ref{Ff})
with $\beta=1$ could hold on large scales.
The problem with this is the extra term $\exp(-2\chi/\sqrt{\alpha})\st{4}{GB}$ which occurs
in the fifth dimension.
The fifth dimension is usually taken to be free of matter so that there is little possibility
of a five dimensional stress cancelling it.
There are surface tension terms in the fourth dimension,
these are terms involving products but not higher derivatives of the stress,
they 'nearly' cancel out the product terms of the Bach tensor (\ref{HX}),
but do not seem to be connected to the Gauss-Bonnet field equations (\ref{Ff}) with $\beta=1$.
Dilatonic terms are likely to contribute to the four dimensional part of the field equation,
and these terms might explain galactic rotation curves \cite{mdrst},
but again dilatonic terms are unlikely to cancel
the extra term in the fifth dimension of (\ref{Ff}).

\end{document}